\documentclass[12pt]{iopart}

\usepackage{epsf}
\usepackage{epsfig}
\usepackage{graphicx}
\usepackage{xcolor}
\usepackage{bm}
\usepackage{tabularx}

\usepackage{amsfonts,amssymb}

\usepackage{float}
\usepackage{array, nicefrac, caption, cases}

\newtheorem{conjecture}{Conjecture}

\newtheorem{proposition}{Proposition}
\newenvironment{proof}{\paragraph{Proof:}}{\hfill$\square$}

\newcommand{\commentout}[1]{}

\newcommand{\x}{x_*}
\newcommand{\M}{\mathbf{M}}
\newcommand{\Mb}{\overline{\mathbf{M}}}
\newcommand{\I}{\mathbf{I}}
\newcommand{\F}{\mathbf{F}}
\newcommand{\Q}{\mathbf{Q}}
\newcommand{\A}{\mathbf{A}}
\newcommand{\U}{\mathbf{U}}
\newcommand{\am}{a}

\newcommand{\binom}[2]{\Bigl(\begin{array}{@{}c@{}}#1\\#2\end{array}\Bigr)}

\begin{document}

\frenchspacing

\title{Flowers of immortality}
\author{Thomas Fink and Yang-Hui He}
\address{
London Institute for Mathematical Sciences, Royal Institution, 21 Albermarle St, London W1S 4BS, UK 
}

\date{\today}

\begin{abstract}
There has been a recent surge of interest in what causes aging.
This has been matched by unprecedented research investment in the field from tech companies.
But, despite considerable effort from a broad range of researchers, we do not have a rigorous mathematical theory of programmed aging.
To address this, we recently derived a mortality equation that governs the transition matrix of an evolving population with a given maximum age.
Here, we characterize the spectrum of eigenvalues of the solution to this equation.
The eigenvalues fall into two classes.
The complex and negative real eigenvalues, which we call the flower, 
are always contained in the unit circle in the complex plane.
They play a negligible role in controlling the dynamics of an aging population.
The positive real eigenvalues, which we call the stem, are the only eigenvalues which can lie outside the unit circle.
They control the most important properties of the dynamics.
In particular, the spectral radius increases with the maximum allowed age. 
This suggests that programmed aging confers no advantage in a constant environment.
However, the spectral gap, which governs the rate of convergence to equilibrium, decreases with the maximum allowed age.
This opens the door to an evolutionary advantage in a changing environment.
\end{abstract}

\maketitle
\vspace*{-15pt}

\commentout{
\section{Introduction} 
\noindent
One of the great mysteries in the life sciences is the universality of aging.
The canonical explanation is that aging is the result of genetic breakdowns in how life processes information.
This is because the selective pressure to combat these breakdowns decreases with with expected total future offspring.

Biological life is at its heart one instance of a self-replicating machine, and we have little understanding, from a thermodynamics point of view, why such machines should arise. At an abstract level, we simply don't understand what are the fundamental constraints on how and why life should 
}
\noindent
\section{Introduction}
\subsection{The mortality equation}
Recently, we derived a simple mortality matrix equation that governs the transition matrix $\Q$ of an evolving population with maximum age $\am$ \cite{Fink22}.
It is
\begin{equation}
\Q^{\am}(\I + \M \F - \Q) = \M \F,
	\label{MatrixMain1}
\end{equation}
where all of the matrices are $2^n \times 2^n$ and $\F$ is a diagonal matrix with the genotype fitnesses along the diagonal. 
The mutation matrix $\M$ satisfies 
\begin{equation}
\M_n = \Mb_n/n,
\end{equation}
where $\Mb$ is defined recursively in block form:
\begin{equation}
\Mb_{n+1} =
\left(
\begin{array}{cc}
  \Mb_n & \I_n    \\ 
   \I_n & \Mb_n     
\end{array}
\right)\!, 
\quad \!
\Mb_1 = \left(
\begin{array}{cc}
	0 & 1  \\
 	1 & 0    
\end{array}\right)\!,
\label{BN}
\end{equation}
with $\I_n$ the $2^n\times 2^n$ identity matrix.
The only difference between $\M$ and $\Mb$ is that $\M$ is a stochastic matrix, in that its rows and columns sum to 1.
\\ \indent
Eq. (\ref{MatrixMain1}) is actually a more compact version of  
\begin{equation}
    \Q^{\am} = \M\F (\I + \Q + \ldots + \Q^{\am-1}),
    \label{MatrixMain2}
\end{equation}
which is the actual governing equation.
Notice that while eq. (\ref{MatrixMain1}) has degree $\am+1$, 
it can be reduced to degree $\am$ by dividing through by $\Q - \I$ to give eq. (\ref{MatrixMain2}).
The solution $\Q = \I$ to eq. (\ref{MatrixMain1}) is spurious, and we always disregard it.
For the derivation of eq. (\ref{MatrixMain2}) and (\ref{MatrixMain1}), see \cite{Fink22}. 
\\ \indent
The matrix $\Q = \M\F$ governs the evolution of a population with maximum age $a = 1$.
It has $2^n$ eigenvalues, some of which may be degenerate.
They are all real and lie in the range $[-1,1]$.

By the Cayley Hamilton theorem, the eigenvalues of $\Q$ have the same functional relation to the eigenvalues of $\M \F$ as the matrix $\Q$ does to the matrix $\M \F$.
For each eigenvalue $\lambda$ of $\M \F$, we obtain a family of $a$ eigenvalues belonging to $\Q$.
This family is generated by the analogue of eq. (\ref{MatrixMain1}), but for numbers rather than matrices:
\begin{equation}
	\mu^a (1 + \lambda - \mu) = \lambda.
	\label{EigenvalueMain1}
\end{equation}
Just as eq. (\ref{MatrixMain1}) is more the compact version of eq. (\ref{MatrixMain2}), albeit with a spurious root at $\Q = \I$, eq. (\ref{EigenvalueMain1}) is a more compact version of
\begin{equation}
   \mu^a = \lambda(1+\mu+\mu^2+\ldots+\mu^{a-1}),
    \label{EigenvalueMain2}
\end{equation}
with a similarly spurious root at $\mu = 1$, which we always disregard.
Eq. (\ref{EigenvalueMain2}) has $a$ solutions, and thus there are $a \, 2^n$ eigenvalues of the matrix $\Q$. The goal of this paper is to characterize them, and thereby better understand the dynamics of an aging population.

Notice that while the eigenvalues $\lambda$ are real, the eigenvalues $\mu$ can be complex.
Whereas a real eigenvalue corresponds to exponential growth or decay of the component of the population that projects along the associated eigenvector, a complex eigenvalue corresponds to oscillatory behavior, with the overall growth or decay set by its magnitude.  
\begin{figure}[p!]
\begin{center}
\includegraphics[width=1\textwidth]{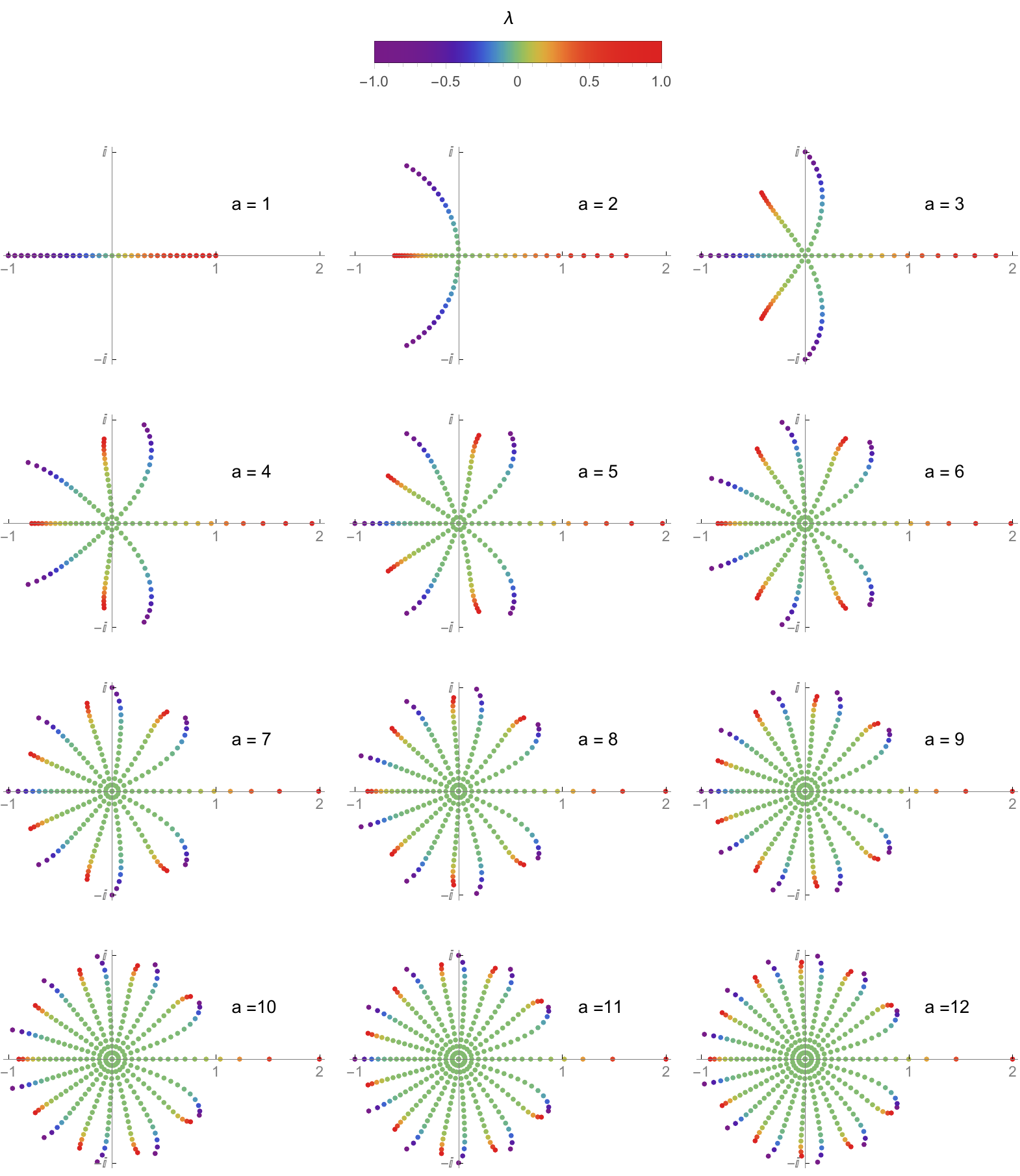}
\caption{\small
	\textbf{All possible eigenvalues of the mortality equation.}
    These are plotted in the complex plane for different values of the maximum allowed age $a$.
    For any given eigenvalue $\lambda \in [-1,1]$ of $\M\F$, there are $a$ eigenvalues of $\Q$, which are plotted with the same color.
    These are the roots of eq. (\ref{EigenvalueMain1}).
    (The curves should be thought of as continuous rather than discrete.)
    The complex and negative real eigenvalues, which we call the flower, are always contained in the unit circle, and play a minor role in the population dynamics.
    The positive real eigenvalues, which we call the stem, dominate the dynamics.
    Of the $a$ eigenvalues associated with a given $\lambda$, the one with the largest magnitude is always the positive real one, and only it can lie outside the unit circle. 
}
\end{center}
\label{RootsPlot}
\end{figure}
\subsection{Summary of results}
In this paper, we do four things, which correspond to the following four sections.
\\ \indent
In the section 2, we study the complex and negative real eigenvalues of the solution $\Q$ to the mortality equation (\ref{MatrixMain1}). 
Our approach is make no assumptions about the fitness $\F$, and therefore about the $2^n$ eigenvalues $\lambda \in [-1,1]$ of $\M\F$.
Rather it is to characterize, for any given $\lambda$,
the family of $a$ eigenvalues satisfying eq. (\ref{EigenvalueMain1}).
The complex and negative real eigenvalues are always contained in the unit circle, which we call the flower (Fig. 1).
These play a negligible role in controlling the dynamics of an aging population.
\\ \indent
In the section 3, we study the real eigenvalues of $\Q$, which we call the stem.
Of the $a$ eigenvalues associated with a given $\lambda$, only one is positive and real, which we call $\rho$, and it has the largest magnitude (Fig. 1).
Only these stem eigenvalues can lie outside the unit circle.
They control most of the important properties of the dynamics of an aging population.
\\ \indent
We study properties of the stem in section 4.
We prove that the stem eigenvalues $\rho$ increase with $\lambda$ and are convex with $\lambda$,
and increase with the maximum age $a$.
The latter is important because it means that, in a fixed environment, programmed aging confers no evolutionary benefit.
We show that the spectral gap of the eigenvalues of $\Q$ increases as $a$ decreases.
This is important because the rate of convergence to equilibrium is controlled by the spectral gap \cite{Pinsky05}.
\\ \indent
In section 5, we test our predictions by calculating the actual eigenvalues for two different fitness functions: uniform fitness and Hamming fitness.
Both are plotted in Fig. 3 and perfectly agree with our predictions.
We conclude with a discussion, where we describe the implications of our results on programmed aging.
\section{The flower: complex and negative real eigenvalues}
In this section we consider the complex and negative real eigenvalues of the solution of the mortality equation.
We call this the flower, because of the petals that appear as the maximum age $a$ increases (Fig. 1).
We show that all these eigenvalues are tame, in the sense that they are contained inside the unit circle on the complex plane.

Our protagonist is the polynomial from eqs. (\ref{EigenvalueMain1}) and (\ref{EigenvalueMain2}), 
recalling that the more compact $P = (\mu-1) Q$ has an $(a+1)$-th spurious root at $\mu=1$:
\begin{eqnarray}
    P(\mu) &:=& - \lambda + (1 + \lambda) \mu^a - \mu^{a+1} ,
    \label{EigenvalueMainP} \\
    Q(\mu) &:=& \lambda(1+\mu+\mu^2+\ldots+\mu^{a-1}) - \mu^a .
    \label{EigenvalueMainQ}
\end{eqnarray}
This canonical form makes it easier to pick off the coefficients in what follows.
We first show that the roots of $Q$ satisfy
\begin{numcases}{}
    \lambda \in [-1,0)  & \quad \mbox{all $a$ eigenvalues are inside the unit circle,}  \label{ComplexRoots1} \\
    \lambda \in (0,1/a] & \quad \mbox{all $a$ eigenvalues are inside the unit circle,} \label{ComplexRoots2} \\
    \lambda \in (1/a,1] & \quad \mbox{$a-1$ eigenvalues are inside the unit circle and one is outside it.} \label{ComplexRoots3}
\end{numcases}
To be clear, when we say inside the unit circle, we include cases where it is on the edge.
Examples are shown in Fig. 1, where the $a$ roots associated with a given $\lambda$ have the same color.
We prove this in three parts, in increasing order of difficulty.

To prove (\ref{ComplexRoots2}), we apply the classic Lagrange Theorem (see, e.g., \cite{polybook}) on bounds of roots: for a polynomial $a_0 + a_1x + \ldots +a_n x^n$, all of its roots lie inside a circle of radius 
\begin{equation*}
\mbox{max}\left[1, \sum\limits_{i=0}^{n-1} \left|\frac{a_i}{a_n} \right| \right].
\end{equation*}
Applying Lagrange to $P(\mu)$, the bounding radius is
\begin{equation*}
\mbox{max}\left[1, \sum\limits_{i=0}^{a-1} \left|\frac{\lambda}{-1} \right| \right] =
    \mbox{max}\left[1, a \lambda \right] = 1.
\end{equation*}
So all $a$ eigenvalues are inside the unit circle.

Proving (\ref{ComplexRoots3}) is more subtle.
We make use of the theorem of Rouch\'e \cite{polybook}, which is as follows.
Let $f(x)$ and $g(x)$ be holomorphic functions inside a region $K \subset \mathbb{C}$ with simple, closed boundary $\partial K$. 
Then if $|g(x)| < |f(x)|$ on $\partial K$, $f$ and $f + g$ have the same number of zeroes inside $K$.
Let us take
\begin{eqnarray}
    \nonumber
    f(\mu) := (\lambda+1) \mu^a - \lambda , \\
    \label{fg}
    g(\mu) := -\mu^{a+1} ,
\end{eqnarray} 
both clearly holomorphic and $f + g = P$.
Next, let $\partial K$ be the unit circle and $K$ be the disk enclosed.
Then $f$ has $a$ roots inside $K$ because its roots are equi-distributed on the circle of radius $\frac{\lambda}{\lambda+1} < 1$.
On $\partial K$, $|g| = 1$ and
$$
    |f| =  \left| (\lambda+1) \mu^a - \lambda \right| >
    \left| (\lambda+1) |\mu^a| - |\lambda| \right| = \left| (\lambda+1) - \lambda \right| = 1,
$$ 
where we used the strict reverse triangle inequality.
So $|g| < |f|$, and therefore $P$ has exactly $a$ eigenvalues in the unit circle, one of which we know is $\mu=1$, which we disregard.
But since $P$ has degree $a+1$, it has $a+1$ roots.
We have accounted for $a$ of them, so the last must be outside the unit circle.

Finally, to prove (\ref{ComplexRoots1}), we modify our Rouch\'e to argument show that all roots lie in the unit circle.
We switch (\ref{fg}) and take 
$f(\mu) := -\mu^{a+1}$ and
$g(\mu) := (1+\lambda) \mu^a - \lambda$,
so $f + g = P$ still.
On the unit circle, $|f| = 1$ while $|g| = |(1+\lambda) \mu^a - \lambda| \leq |1+\lambda| + |\lambda|$ by the triangle inequality. 
However, equality is only when $(1+\lambda) \mu^a = \lambda$, which is not possible on the unit circle except when $\lambda = -\frac12$, a case which we will address separately.
In all other cases, recalling that $\lambda < 0$, $|g| < |1+\lambda| + |\lambda| = 1+\lambda - \lambda = 1 = |f|$ on the unit circle and clearly $f$ has $a+1$ roots inside. 
Thus by Rouch\'e, $P$ has $a+1$ roots inside the unit circle, one of the which is the spurious root on the edge.
For $\lambda=-1/2$, take $f = 1/2 - \mu^{a+1}$ and $g = 1/2$ so that $f+g = P$.
Now, on the unit circle 
\begin{equation*}
|f| = \left|1/2 - \mu^{a+1} \right| > \left| 1/2 - |\mu^{a+1}| \right| = |1/2 - 1|=1/2,
\end{equation*} 
where we have used the negative reverse triangle inequality. 
So $|g| < |f|$, with $f$ having $a+1$ roots, and we have the same situation.

We mention in passing that this property of polynomials---where one root lies outside the unit disk and all others lie inside---is a well-studied subject of Galois theory.
An algebraic integer outside the unit circle, whose Galois conjugates all lie within, is called a Pisot number \cite{pisot}.
\section{The stem: positive real eigenvalues}
Having described the locations of the eigenvalues with respect to the unit circle, we investigate the real roots.
We call the positive real roots the stem, because they lead to the flower with its $a$ petals (Fig. 1).
As we will see, the stem determines most properties of the dynamics of an aging population that we might be interested in.
\subsection{The positive root is the dominant root}
We start by showing that for $\lambda<0$ there are no positive roots.
We appeal to Descartes' Rule of Signs \cite{polybook}:
for a polynomial $p(x)$ with real coefficients, 
in which the non-zero coefficients are ordered by ascending exponents in $x$, 
the number of positive roots of $p(x)$ is either the number of sign changes between consecutive coefficients, or is less than it by an even number. 
Similarly, the number of negative roots is counted using $p(-x)$.
With $\lambda$ negative in eq. (\ref{EigenvalueMainQ}), all coefficients of $Q$ have the same sign, so there can be no positive roots.

However, for $\lambda > 0$, the eigenvalue with the largest magnitude, which we
call $\rho$, is always the only positive real one.
This follows from Perron-Frobenius \cite{polybook}: for a positive matrix $\A$, there is a positive real eigenvalue whose magnitude exceeds all others.
The $a \times a$ matrix here is 
\begin{equation}
    \A = \left( \begin{array}{cccccc}
       \lambda  & \lambda & \lambda & \ldots & \lambda & \lambda\\
       1  & 0 & 0 & \ldots & 0 & 0\\
       0  & 1 & 0 & \ldots & 0 & 0\\
       \vdots & \vdots &\vdots &\vdots &\vdots & \vdots \\
       0  & 0 & 0 & \ldots & 1 & 0\\
    \end{array} \right),
\end{equation}
for which (\ref{EigenvalueMainQ}) is the characteristic polynomial.

We now rewrite eqs. (\ref{MatrixMain1}) and (\ref{MatrixMain2}) in terms of $\rho$ instead of $\mu$,
\begin{eqnarray}
	\rho^a (1 + \lambda - \rho) = \lambda,	            \label{MaxEigenvalueMain1} \\
	\rho^a = \lambda(1+\rho+\rho^2+\ldots+\rho^{a-1}),  \label{MaxEigenvalueMain2}
\end{eqnarray}
always bearing in mind that $\rho$ is positive and real.
From the previous section, we know that $\rho$ satisfies
\begin{equation} \label{prop:root}
\left\{
\begin{array}{ll}
    \lambda \in [-1,0)  & \quad \rho < 1,   \\
    \lambda \in (0,1/a] & \quad \rho = 1,   \\
    \lambda \in (1/a,1] & \quad \rho > 1.  
    \end{array}
    \right.
\end{equation}

We will refine these bounds in the next section.
But before we do, we show that, furthermore,
when is $a$ odd, there are no other real roots, 
but when is $a$ even, there is one other real root, which is negative.
We can see this geometrically by rearranging (\ref{EigenvalueMainP}) into $\mu^a = \frac{\lambda}{1+\lambda-\mu}$.
Then the roots occur at the intersection of a monomial and a shifted hyperbola.
For a more systematic approach, we again use Descartes' Rule of Signs.
\begin{figure}[b!]
\begin{center}
\includegraphics[width=0.67\textwidth]{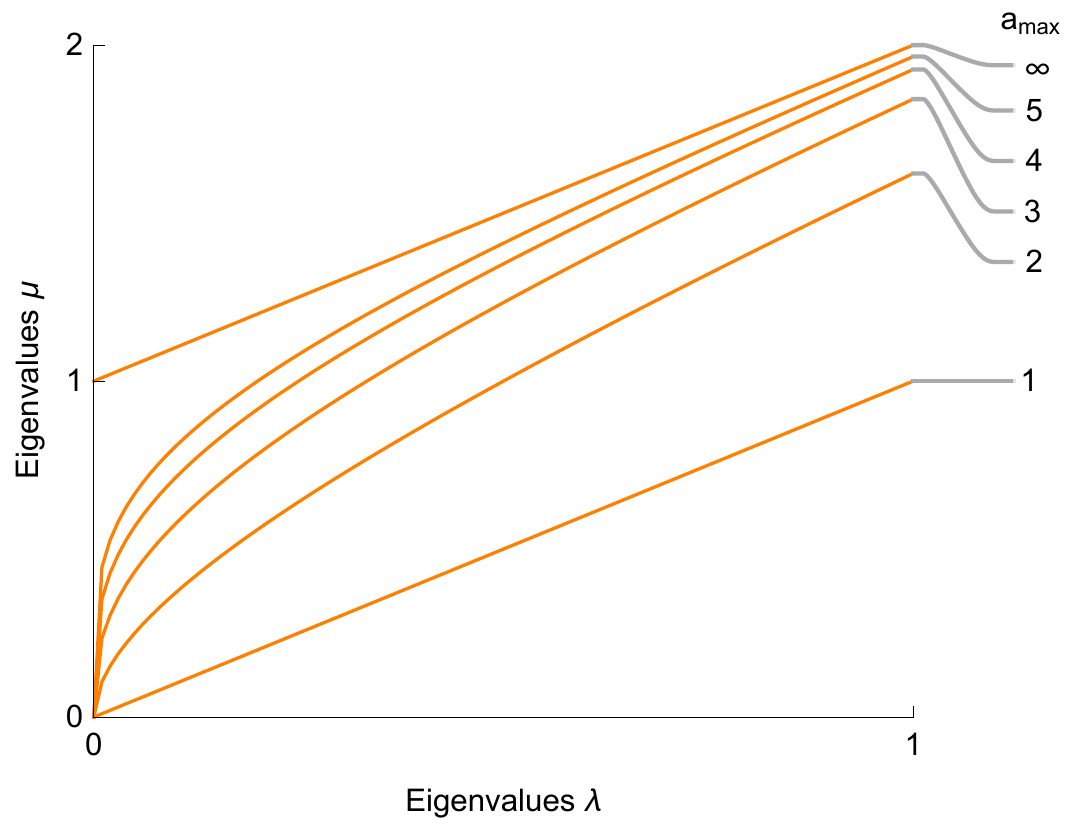}
\caption{\small
	\textbf{Eigenvalues for different maximum ages.}
	For any eigenvalue $\lambda$ of the transition matrix $\M\F$ ($a=1$), the eigenvalue $\mu$ of the transition matrix $\Q$ for higher $a$ is given by jumping up to the appropriate line.
    	The fitness $\F$ alone determines the horizontal placement of the eigenvalues, but $a$ alone determines their vertical placement.
}
\label{EigenvalueTransform}
\end{center}
\end{figure}
For $\lambda \in (0,1]$, $Q(\mu)$ has coefficients 
$\lambda, \lambda, \ldots, -1$, 
so there is a single sign change, accounting for the sole positive root $\rho$.
Next, consider $P(-\mu)$, for which the  coefficients are
$-\lambda, (1+\lambda) (-1)^a, -(-1)^{a+1}$.
So if $a$ is odd, there is no sign change and hence no negative root, but  
if $a$ is even, there is a single sign change and a single negative root. 
\subsection{Bounds on the positive roots}
We now develop some useful bounds on the magnitude of $\rho$, which refine (\ref{prop:root}):
\begin{equation} \label{Bounds}
\left\{
\begin{array}{ll}
        \lambda \in (0,1/a)     & \quad \lambda \,<\, \lambda^{1/a} \,<\, \rho \,<\, (a\lambda)^{1/a} \,<\, \lambda + \frac{a-1}{a} \,<\, 1, \\
        \lambda = 1/a           & \quad \rho = 1, \\
        \lambda \in (1/a, 1]    & \quad 1 \,<\, (a\lambda)^{1/a} \,<\, \lambda + \frac{a-1}{a} \,<\, \rho \,<\, 1 + \lambda.
    \end{array}
    \right.
\end{equation}

Let's first establish that $\rho>\lambda$, which we will use below.
To do so, we set $\rho=\lambda+\epsilon$, and show $\epsilon$ must be positive.
Substituting $\lambda = \rho - \epsilon$ into (\ref{MaxEigenvalueMain1}), we find
$\epsilon = (\rho^a-\rho)/(\rho^a-1)$.
For $\lambda<1/a$, both numerator and denominator are negative, and for $\lambda>1/a$, both are positive.
So $\epsilon$ is positive and $\rho > \lambda$.

Now let's establish that $\rho<1+\lambda$.
To do so, we set $\rho = 1 + \lambda - \epsilon$, and show $\epsilon$ must be positive.
Substituting $\lambda = \rho - 1 + \epsilon$ in (\ref{MaxEigenvalueMain1}), we find
$\epsilon = (\rho-1)/(\rho^a-1)$.
For $\lambda<1/a$, both terms are negative, and for $\lambda>1/a$, both are positive. 
So $\epsilon$ is positive and $\rho < 1 + \lambda$.

Turning to eq. (\ref{MaxEigenvalueMain2}), we see that $\rho^a < a \lambda$ for $\lambda < 1/a$, since $\rho<1$ and each of the terms $1+\rho+\rho^2+\ldots$ can be at most 1.
Likewise, $\rho^a > a \lambda$ for $\lambda > 1/a$, since $\rho>1$.
Substituting these bounds on $\rho^a$ into $\rho^a = \lambda/(1+\lambda-\rho)$ gives 
$\rho < \lambda + \frac{a-1}{a}$ for $\lambda < 1/a$ and
$\rho > \lambda + \frac{a-1}{a}$ for $\lambda > 1/a$.
From (\ref{MaxEigenvalueMain1}), $\rho^a = \frac{\lambda}{1+\lambda-\rho} > \lambda$ because $\rho>\lambda$. 
So $\rho > \lambda^{1/a}$ for all $\lambda$.

It only remains to show that $(a\lambda)^{1/a} < \lambda + \frac{a-1}{a}$ for all $\lambda$, 
apart from $1/a$ where the inequality is an equality.
With $x = a \lambda$, this can be written $a^a x < (x+a-1)^a$.
For $a>1$, the left side is linear in $x$, whereas the right side is convex in $x$. We can see that the left side is the tangent below the right side that touches at $x=1$, since both sides evaluate to $a^a$ and have derivative $a^a$ at $x=1$.
\section{Properties of the stem}
Understanding the stem, which is the key to the dynamics, amounts to understanding the only positive real solution to eqs. (\ref{MaxEigenvalueMain1}) and (\ref{MaxEigenvalueMain2}).
We focus on this in this section, where we now only consider $\lambda \in (0,1]$.
\subsection{The dominant root $\rho$ is increasing with $\lambda$}
To show that the dominant root $\rho$ is increasing with $\lambda$, we need to show that the first derivative with respect to $\lambda$ is positive.
Implicitly differentiating eq. (\ref{MaxEigenvalueMain1}) with respect to $\lambda$ and solving for $\rho' = d\rho/d\lambda$ gives
\begin{equation}
    \rho' = \frac{\rho (\rho-1)}{\lambda D},
    \label{FirstDerivative}
\end{equation}
where $D$, which stands for denominator, is
\begin{equation}
    D = \rho - a(1+\lambda-\rho) = \rho - a \lambda / \rho^a.
    \label{Ddef}
\end{equation}
Since $\rho$ and $\lambda$ are positive, the condition on $\rho'$ being positive is
\begin{equation}
    \frac{\rho-1}{D} > 0.
    \label{IncCond}
\end{equation}
\indent
Inserting the $(a\lambda)^{1/a}$ bounds from (\ref{Bounds}) into (\ref{Ddef}), we find
\begin{equation*}
\left\{
\begin{array}{ccl}
        \lambda < 1/a &:& D < (a\lambda)^{1/a} - 1 < 0, \\
        \lambda = 1/a &:& D = 0, \\
        \lambda > 1/a &:& D > (a\lambda)^{1/a} - 1 > 0.
    \end{array}
    \right.
\end{equation*}
For $\lambda < 1/a$, both $D$ and $\rho-1$ are negative in (\ref{IncCond}), while they are both positive 
for $\lambda > 1/a$.
Thus we see that $\rho'$ is everywhere positive.
\subsection{The dominant root $\rho$ is convex with $\lambda$}
To show that $\rho$ is convex, we simply need to show that that the second derivative is negative.
We can write $\rho''$ as
\begin{equation*}
    \rho'' = \frac{a \rho (\rho-1) (1+\lambda-\rho) (1-a\lambda+D)}{\lambda^2 D^3},
\end{equation*}
where $D$ is defined in (\ref{Ddef}).
Dropping terms that are always positive, the condition on $\rho''$ being negative is
\begin{equation*}
    \frac{(\rho-1)(1-a\lambda+D)}{D} < 0.
\end{equation*}

For $\lambda < 1/a$, $D$ and $\rho-1$ are both negative, and $1-a\lambda+D < (a\lambda)^{1/a} - a\lambda < 0$. So $\rho'' < 0$.

For $\lambda > 1/a$, $D$ and $\rho - 1$ are both positive.
We just need to show that 
\begin{equation}\label{F}
F := 1-a\lambda+D = (a+1)u - 2a \lambda - a + 1 < 0. 
\end{equation}
This is more subtle.
Our approach is to note that $F$ is 0 at $\lambda=1/a$ (since $u=1$ there), and falls off to the right.
To see this, let's compute the derivative of $F$ with respective to $\lambda$, remembering that $u$ implicitly depends on $\lambda$. We find that
\begin{equation}
    F' = \frac{(a+1)\rho(\rho-1)-2a\lambda D}{\lambda D} <
    -\frac{\lambda (1+\lambda) (a-1)}{\lambda D}
\end{equation}
where we used the fact that $\rho < 1 + \lambda$.
All of the terms in the right side are always positive, apart from $D$.
For $\lambda>1/a$, $D>0$, so $F' < 0$, and thus for $\lambda \in (1/a, 1]$, $F$ is decreasing from 0.
Thus we see that $F<0$ $p''$ is everywhere negative.
\subsection{The dominant root $\rho$ increases with $a$}
Our approach to showing that $\rho$ increases with the maximum age $a$ is to treat $a$ as though it were a continuous parameter instead of an integer, differentiate $\rho$ with respect to it, and show that the result is positive.
This approach is justified by analytic continuation---the defining equation is clearly analytic in all variables, and in $a > 0$ in particular.
Implicitly differentiating eq. (\ref{MaxEigenvalueMain1}) with respect to $a$ and solving for $d\rho/da$ gives
\begin{equation}
    \frac{d\rho}{da} = \frac{(1+\lambda-\rho) \rho \ln \rho}{D},
    \label{RadiusMain}
\end{equation}
where $D$ is defined in (\ref{Ddef}).
Dropping terms that are always positive, the condition on $d\rho/da$ being positive is
\begin{equation}
    \frac{\ln \rho}{D} > 0.
    \label{NN}
\end{equation}
Since $\rho<1$ and $D<0$ for $\lambda < 1/a$,
and $\rho>1$ and $D>0$ for $\lambda > 1/a$,
$d\rho/da$ is always positive.
\subsection{Spectral gap decreases with age}
We now show that the spectral gap decreases with the maximum age $a$.
Let $\lambda_1$ and $\lambda_2$ be the first and second eigenvalues of the matrix $\M \F$, which is the transition matrix for $a=1$. 
The corresponding eigenvalues for $a>1$ are given by eq. (\ref{MaxEigenvalueMain1}).
Let $b>a$ be an integer. 
Then our condition for the spectral gap to decrease with $a$ is
\begin{equation}
\frac{\rho(\lambda_1,a)}{\rho(\lambda_2,a)} > \frac{\rho(\lambda_1,b)}{\rho(\lambda_2,b)}
\end{equation}
With $\lambda_2 = \lambda$ and $\lambda_1 = \lambda + \Delta \lambda$, this becomes
$
    \frac{\rho(\lambda + \Delta \lambda, a)}{\rho(\lambda,a)} > \frac{\rho(\lambda + \Delta \lambda, b)}{\rho(\lambda,b)}
$.
In the limit of $\Delta \lambda \rightarrow 0$, $\rho(\lambda + \Delta \lambda,a) = \rho(\lambda,a) + \Delta \rho(\lambda,a)$, and we have
$
\frac{\Delta \rho(\lambda,a)}{\rho(\lambda,a)} > \frac{\Delta \rho(\lambda, b)}{\rho(\lambda,b)}$.
Dividing both sides by $\Delta \lambda$, and with $\rho' = d\rho/d\lambda$, this becomes
\begin{equation}
	\frac{\rho'(\lambda,a)}{\rho(\lambda,a)} 
	>
	\frac{\rho'(\lambda,b)}{\rho(\lambda,b)}.
\end{equation}

Our approach is to show that $\rho'/\rho$ decreases with $a$.
As before, we will analytically continue and consider the derivative $d/da$.
From (\ref{FirstDerivative}),
\begin{equation*}
    \frac{\rho'}{\rho} = \frac{\rho-1}{\lambda D}.
\end{equation*}
Then
\begin{equation*}
\frac{d}{da} \frac{\rho'}{\rho} = \frac{(\rho-1)(1+\lambda-\rho) - (a \lambda -1) d\rho/da}{\lambda D^2}.
\end{equation*}
Substituting in $d\rho/da$ from (\ref{RadiusMain}) yields
\begin{eqnarray*}
\frac{d}{da} \frac{\rho'}{\rho}   
                            &=& \frac{(\rho-1)D - (a \lambda -1) u \ln \rho}{\rho^a D^3}. 
\end{eqnarray*}
We want to show that this is always negative.
Dropping positive terms, it's sufficient to show that 
\begin{equation}
\frac{(\rho-1)D - (a \lambda -1) \rho \ln \rho}{D} < 0.
\label{PA}
\end{equation}
We will show this first for $\lambda < 1/a$ and then for $\lambda > 1/a$.
In doing so, we make use of the well-known log-inequality, that for $z>0$,
\begin{equation*}
    1 - \frac{1}{z} \leq \ln(z) \leq z - 1,
\end{equation*}
with equality at $z=1$ and inequalities elsewhere.
\\ \indent
First, consider $\lambda>1/a$. 
Then $\rho>1$ and $D>0$.
We need only show that the numerator of (\ref{PA}) is negative, that is, $N := (\rho-1)D - (a \lambda -1) \rho \ln \rho < 0$.
Using the left side of the $\ln z$ inequality, this is implied by
\begin{equation}
(\rho-1)(D + 1 - a \lambda) = (\rho-1) F < 0,
\end{equation}
where $F$ was defined in (\ref{F}).
Since we know $F<0$ for all $\lambda$, $N < 0$, and $N/D < 0$.

Second, consider $\lambda<1/a$. 
Then $\rho<1$ and $D<0$.
Now we need to show that $N > 0$.
Again using the left side of the $\ln z$ inequality, this is implied by
\begin{equation}
    (\rho-1) F > 0.
\end{equation}
Since $F<0$ for all $\lambda$, and $\rho-1$ is now negative, $N > 0$, and $N/D < 0$

Thus $\frac{d}{da} \frac{\rho'}{\rho}$ is always negative and the spectral gap decreases with maximum age $a$.
\begin{figure}[p!]
\begin{center}
\includegraphics[width=1\textwidth]{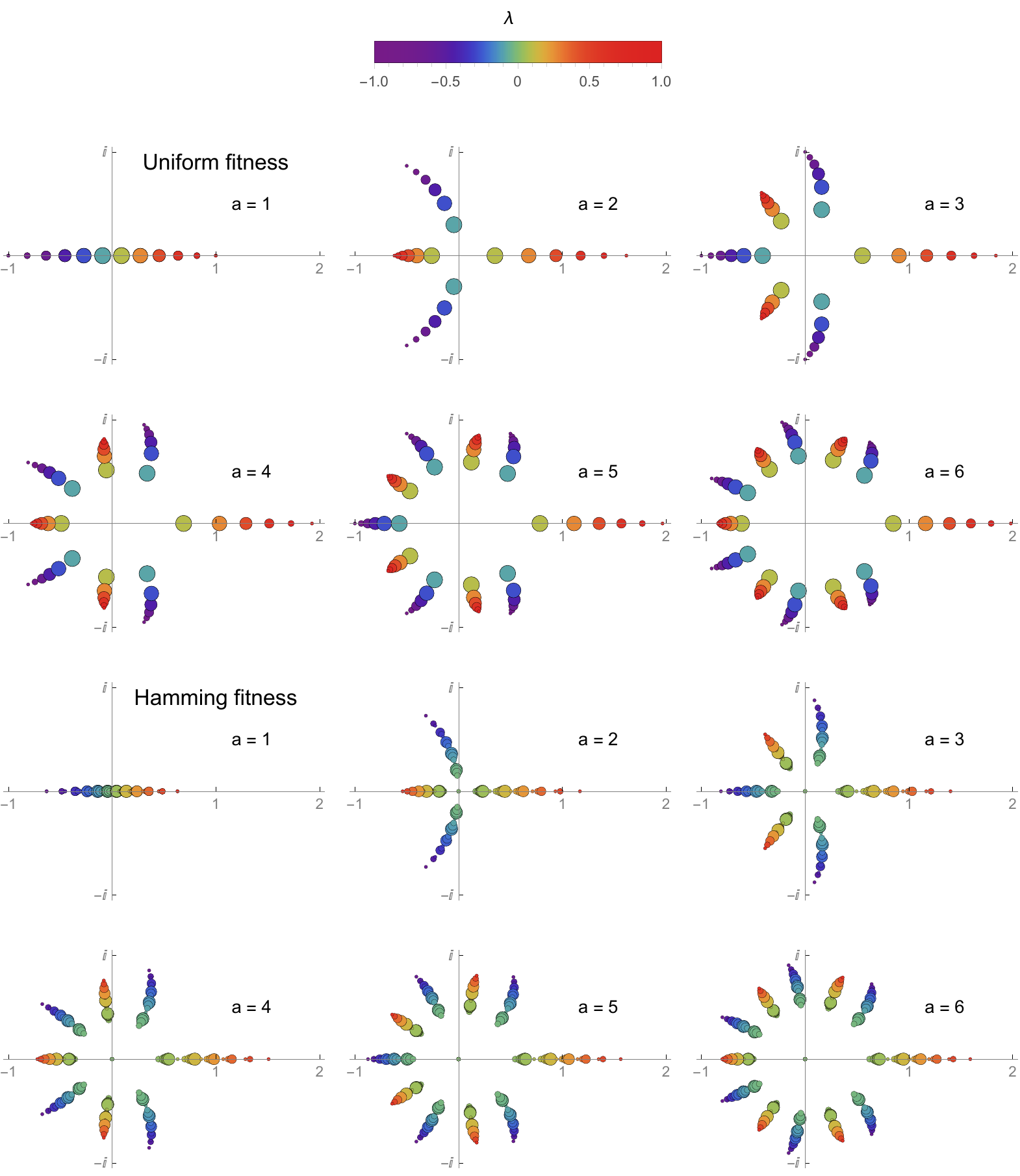}
\caption{\small
	\textbf{Actual eigenvalues of the mortality equation, for uniform fitness and Hamming fitness.}
    Here we show all $a \, 2^n$  eigenvalues of the mortality equation, for maximum age $a=1$ to $a=6$, and genome length $n=11$.
	The degeneracy is proportional to the square of the area of the circles 
	(to avoid overcrowding of the points).
	Note that in both cases the plots superimpose onto the plots in Fig. 1.
}
\end{center}
\label{RootsPlot}
\end{figure}
\section{Testing our predictions and discussion}
\subsection{Testing our predictions}
To test our prediction of the spectrum of eigenvalues of the transition matrix $\Q$, 
we numerically calculated the actual spectrum for two different fitness functions $\F$.
In both cases we set the genome length to $n = 11$.
First, we considered constant fitness, in which every genotype has fitness 1: $\F$ is just the identity matrix $\I$. 
Constant fitness diffuses the population over the hypercube, smoothing it out towards a uniformly distributed population.
The result is shown in the top half of Fig. 3. 

Second, we considered the Hamming fitness. 
This corresponds to a natural notion of distance: 
the number of edges on the hypercube that must be traversed to get from one corner to another.
The Hamming fitness is one minus $1/n$ times the Hamming distance between some genotype $g$ and the optimal genotype $\tilde{g}$, that is, 
the fraction of bits where $g$ and $\tilde{g}$ match.
The result is shown in the bottom half of Fig. 3. 

In both cases, the eigenvalues for the real systems perfectly superimpose on our plot of all possible eigenvalues in Fig. 1.
This confirms our predictions.
\subsection{Discussion}
For an evolving population with maximum age $a$ and an arbitrary fitness function, the population transition matrix $\Q$ is the solution to the mortality equation $\Q^{\am}(\I + \M \F - \Q) = \M \F$, which we derived previously \cite{Fink22}.
The eigenvalues of the matrix $\Q$ govern the behavior of the population over time.
There are a total of $a \, 2^n$ of them, though in practice many of them will be degenerate.
In particular, there is a family of $a$ eigenvalues generated by each eigenvalue $\lambda \in [-1,1]$ of $\M\F$.
\\ \indent
These eigenvalues fall into two classes (Fig. 1).
We call the complex and negative real eigenvalues the flower, and they are all bounded by the unit circle in the complex plane. 
Because of this, these eigenvalues play a negligible role in governing important properties of the population dynamics.
\\ \indent
We call the positive real eigenvalues the stem (Fig 1). 
For the family of eigenvalues associated with a given $\lambda$, the one with the greatest magnitude is always the only positive real one, which we call $\rho$. 
Only these positive eigenvalues can lie outside the unit circle.
\\ \indent
The magnitude of the positive real eigenvalue $\rho$ always dominates, but the extent to which it does so falls into two regimes.
For $\lambda < 1/a$, all of the eigenvalues approximately lie on a circle centered at the origin, 
with $\rho$ slightly farther from the origin than the others.
However, when $\lambda > 1/a$, the dominant eigenvalue $\rho$ starts to break away. 
As $\lambda$ goes from $1/a$ to 1, all of the eigenvalues except $\rho$ approach the unit circle, but never exceed it.
But $\rho$ goes from 1 to the $a$-bonacci constant $\phi_a$, eclipsing the others, where $\phi_2 = 1.62$, $\phi_3 = 1.84$, and so on, up to $\phi_\infty = 2$ \cite{Wolfram96}.
\\ \indent
In a fixed environment, the long-term growth rate of the population is set by the dominant eigenvalue of $\Q$, which is also known as the spectral radius.
Since this increases with $a$, a population with a higher maximum age $a$ grows faster than one with a smaller $a$.
Thus there is no growth rate benefit afforded by aging in a fixed environment.
Mortality is a losing strategy and programmed aging is not favored by natural selection.
\\ \indent
When the two largest eigenvalues $\rho_1$ and $\rho_2$ of $\Q$ are positive and real---which is guaranteed to be the case for sufficiently large $a$---the spectral gap $\rho_1/\rho_2$ increases as the age $a$ decreases.
Since the rate of convergence to equilibrium is controlled by the spectral gap, this opens up the possibility of an evolutionary advantage in a changing environment.
~\\
~\\


\begin{thebibliography}{99}
\bibitem{Fink22}    T. Fink,        Mortality equation characterizes the dynamics of an aging population,       https://arxiv.org/abs/2208.14915.
\bibitem{Pinsky05}	R. Pinsky,		Spectral gap and rate of convergence to equilibrium for a class of conditioned Brownian motions,	Stoch Proc Appl,		{\bf 115},	875			(2005). 	
\bibitem{Brauer50} 	A. Brauer, 		On algebraic equations with all but one root in the interior of the unit circle,		Amer Math Month {\bf 4}, 250 (1950).
\bibitem{polybook}  Q.~I.~Rahman, G.~Schmeisser, Gerhard    {\it Analytic theory of polynomials}, LMS Lectures {\bf 26}, Oxford University Press (2002).
\bibitem{pisot}     C.~Pisot,                  La r\'epartition modulo 1 et nombres alg\'ebriques,   Ann.~Sc.~Norm.~Super.~Pisa II. Ser.~{\bf 7} 205 (1938).
\bibitem{Wolfram96}   D. A. Wolfram,                  Solving generalized Fibonacci recurrences. 
Fibonacci Quarterly {\bf 36}, 129 (1998)
\end{thebibliography}
\end{document}